\documentclass[prb,twocolumn,superscriptaddress,preprintnumbers,amsmath,amssymb]{revtex4}
\usepackage{hyperref}
\usepackage{cleveref}
\usepackage{gensymb}
\usepackage{graphicx}
\usepackage{dcolumn}
\usepackage{bm}
\usepackage{color}
\newcommand{\be}{\begin{equation}}
\newcommand{\ee}{\end{equation}}
\newcommand{\bea}{\begin{eqnarray}}
\newcommand{\eea}{\end{eqnarray}}
\newcommand{\bp}{\mathbf{p}}
\newcommand{\bk}{\mathbf{k}}
\newcommand{\br}{\mathbf{r}}

\hypersetup{colorlinks=true,linkcolor=blue,citecolor=blue}

\begin{document}

\title{Tunable properties of excitons in double monolayer semiconductor heterostructures}

\author{Luiz G. M. Ten\'orio}\email{luiztenorio@fisica.ufmt.br}
\affiliation{Instituto de F\'isica, Universidade Federal de Mato Grosso, 78060-900, Cuiab\'a, Mato Grosso, Brazil}
\affiliation{Departamento de F\'isica, Instituto Tecnol\'ogico de Aeron\'autica, DCTA, 12228-900 S\~ao Jos\'e dos Campos, Brazil}

\author{Teldo A. S. Pereira}\email{teldo@fisica.ufmt.br}
\affiliation{Instituto de F\'isica, Universidade Federal de Mato Grosso, 78060-900, Cuiab\'a, Mato Grosso, Brazil}

\author{K. Mohseni}\email{kmohseni@ita.br}
\affiliation{Departamento de F\'isica, Instituto Tecnol\'ogico de Aeron\'autica, DCTA, 12228-900 S\~ao Jos\'e dos Campos, Brazil}

\author{T. Frederico}\email{tobias@ita.br}
\affiliation{Departamento de F\'isica, Instituto Tecnol\'ogico de Aeron\'autica, DCTA, 12228-900 S\~ao Jos\'e dos Campos, Brazil}

\author{M.~R.~Hadizadeh}\email{mhadizadeh@centralstate.edu}
\affiliation{College of Engineering, Science, Technology and Agriculture, Central State University, Wilberforce, OH, 45384, USA} 
\affiliation{Department of Physics and Astronomy, Ohio University, Athens, OH, 45701, USA}

\author{Diego R. da Costa}\email{diego\_rabelo@fisica.ufc.br}
\affiliation{Departamento de F\'isica, Universidade Federal do Cear\'a, Campus do Pici, 60455-900 Fortaleza, Cear\'a, Brazil}

\author{Andr\'e J. Chaves} \email{andrejck@ita.br}
\affiliation{Departamento de F\'isica, Instituto Tecnol\'ogico de Aeron\'autica, DCTA, 12228-900 S\~ao Jos\'e dos Campos, Brazil}

\date{\today}

\begin{abstract}
We studied the exciton properties in double layers of transition metal dichalcogenides (TMDs) with a dielectric spacer between the layers. We developed a method based on an expansion of Chebyshev polynomials to solve the Wannier equation for the exciton. Corrections to the quasiparticle bandgap due to the dielectric environment were also included via the exchange self-energy calculated within a continuum model. We systematically investigated hetero double-layer systems for TMDs with chemical compounds MX$_2$, showing the dependence of the inter- and intralayer excitons binding energies as a function of the spacer width and the dielectric constant. Moreover, we discussed how the exciton energy and its wave function, which includes the effects of the changing bandgap, depend on the geometric system setup.
\end{abstract}
\maketitle

\section{\label{sec_introduction}Introduction}

The wide variety of two-dimensional (2D) materials with different properties has opened up the possibility of atomic scale heterogeneous integration and combination of different layers, thus creating new hybrid structures that exhibit totally new physics and allow unique functionalities. A relevant perspective review paper in 2013 named this mixing of isolated layers into stacked heterostructures as van der Waals heterostructures\cite{geim2013van}. Such layer-stacked junctions have been intensively explored in the past decade, presenting novel optoelectronics and collective quantum phenomena that, in turn, one shows to be a highly tunable material platform to design new high-performance nanoelectronic devices tailored to a specific purpose based on the layers' compounds choice\cite{li2016heterostructures, raja2017Coulomb, zhang2016van, mak2016photonics, chaves2020bandgap}.

A promising research area within optoelectronics in semiconductor 2D materials and its layered structures is related to the fact that they support the formation of excitons -- bound electron-hole pairs -- and excitonic complexes with binding energies more than an order of magnitude greater than conventional semiconductors, i.e., on the order of hundreds of meV, and small Bohr radius in the range of several manometers \cite{berkelbach2013theory, zhang2014direct, he2014tightly, ye2014probing, ugeda2014giant, cheiwchanchamnangij2012quasiparticle, cavalcante2018stark, chaves2021signatures}. It stems from the reduced dimensionality and the associated reduced dielectric screening that, in turn, leads to strong Coulomb interactions between the charge carriers. Consequently, the energy levels are renormalized, the quasiparticle bandgap is modified, and the exciton binding energy can be tuned by changing the environment \cite{bernardi2017optical, kylanpaa2015binding, lamountain2018environmental}. Therefore, an alternative to control the strength of the Coulomb interaction via structural, sizable, and dielectric environment is engineering the van der Waals stacking \cite{lamountain2018environmental, zhang2016van, latini2015excitons, andersen2015dielectric}, and consequently, the interlayer electrostatic coupling between the constituents, leading to a weakening or strengthening of the Coulomb binding by increasing or decreasing the spatial separation between the electron and the hole.

Owing to the interplay between the layer-dependence control and the highly sensitive excitonic effects in van der Waals materials, allowing the existence of a huge amount of different combinations of interlayer and intralayer excitons in homostructures and heterostructures \cite{calman2018indirect, calman2020indirect, ruiz2020theory, viner2021excited}, aligned with numerous different reported techniques to deal with excitonic complexes and even Bose-Einstein condensate of excitons \cite{Wang2019}, motivates further exploration of methods to compute exciton properties given the richness of possibilities to create and control them. 

In this work, we present a simple yet efficient and accurate method, being less computationally demanding than the Bethe-Salpeter framework from first-principles and Monte Carlo approaches and with accurate convergence in comparison with other semi-analytical methodologies based on 2D hydrogenic excitonic basis\cite{quintela2020colloquium, henriques2019optical, gomes2021variational, rodin2014excitons}, to solve the excitonic Wannier equation within the effective mass approximation by using a basis expansion of the eigenstate wave function into the Chebyshev's polynomials. Results for the dependence of the exciton energy levels (binding energies) and associated wave functions on the layer separation and dielectric constant of dielectric spacers are obtained for interlayer and intralayer excitons in different combinations of double-layer transition metal dichalcogenides (TMDs) composing heterostructures.

The paper is organized as follows. In Sec.~\ref{sec.method}, we present the theoretical framework used to solve the excitonic Wannier equation, deriving, in Appendix~\ref{subsec.poisson}, from the Poisson equation for double-layer system separated by a spacer the appropriate intralayer and interlayer electrostatic potential contributions, and in Sec.~\ref{subsec.tcheb} we demonstrate the solution of Wannier equation for excitons by expanding the excitonic wave function in Chebyshev polynomials to obtain the binding energies and wave function in real and momentum spaces. Sec.~\ref{sec.bandgap} is devoted to explaining the procedure to find the bandgap correction for double-layer semiconductors taking into account the found electrostatic interaction and starting from the monolayer bandgap. Results for heterostructures are discussed in Sec.~\ref{sec_results} comparing them with the previously reported results. Finally, in Sec.~\ref{sec.conclusions}, we summarize our main findings.

\section{Methodology}\label{sec.method}

\begin{figure}[t]
\centering
\includegraphics[width=1\columnwidth]{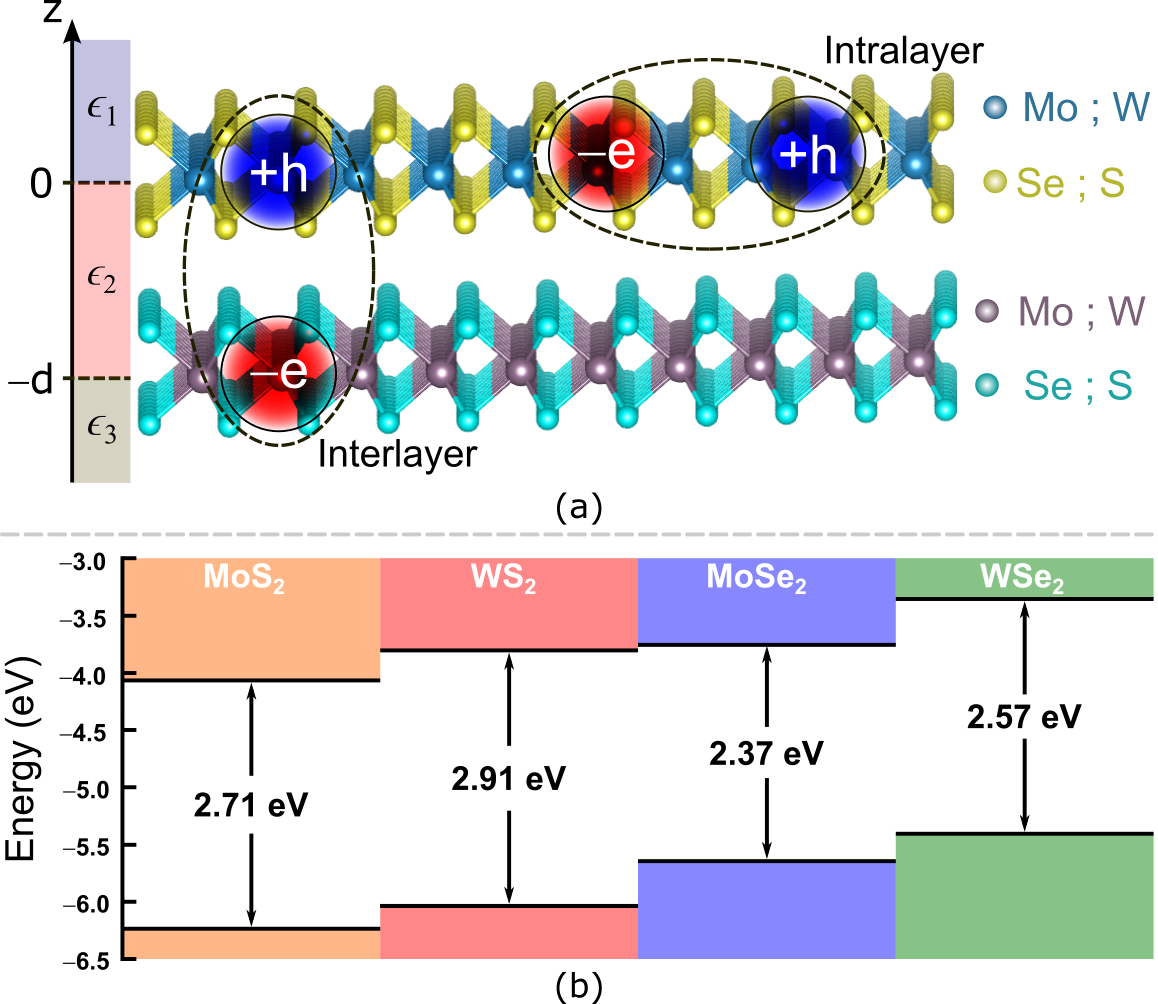}
\caption{(Color online) (a) Schematic illustration of the double layered TMDs, separated by a spacer of dielectric constant $\epsilon_2$ ($-d\leq z\leq 0$) and width $d$, immersed in two materials of dielectric constants $\epsilon_1$ ($z>0$) and $\epsilon_3$ ($z<-d$). This structure sustains both intralayer and interlayer excitons. (b) Band alignment as measured from the vacuum between the four TMDs considered in this work. The bandgap energies and their alignments were obtained from DFT calculations in Ref.~[\onlinecite{Zhang_2016}].}
\label{Fig1}
\end{figure}

We investigate two semiconductor monolayers separated by a spacer with width $d$ and dielectric constant $\epsilon_2$. The substrate ($z<-d$) and superstrate ($z>0$) have dielectric constants $\epsilon_3$ and $\epsilon_1$, respectively, as depicted in Fig.~\ref{Fig1}(a). Here, we consider different TMDs semiconductors represented by the symbol MX$_2$, where $M$ is a metal [molybdenum ($Mo$) or tungsten ($W$)] and $X$ is a chalcogenide [selenium ($Se$) or sulfur ($S$)]. Homo and heterostructures are formed by taking the same or different TMDs in the double-layer system, respectively. In Fig.~\ref{Fig1}(b), we depict the energy gap values for the four investigated TMDs here. Note that the resulting heterobilayers lead to a type II band alignment\cite{Zhang_2016}, that strongly favors the formation of interlayer excitons \cite{latini2017interlayer}. To correctly predict the exciton energies, determined as the difference between the bandgap and the magnitude of the exciton binding energy, we consider the effects of the dielectric geometry on the carrier-carrier interaction as the solution of the corresponding Poisson equation. We use the Wannier equation in the effective mass approximation to calculate the exciton energy, which was proven to coincide with a microscopic model \cite{have2019tmd}. For the bandgap, we use the exchange self-energy\cite{Chaves_2017} within the continuum model.

\begin{figure}[htpb!]
\centering
\includegraphics[width=1\columnwidth]{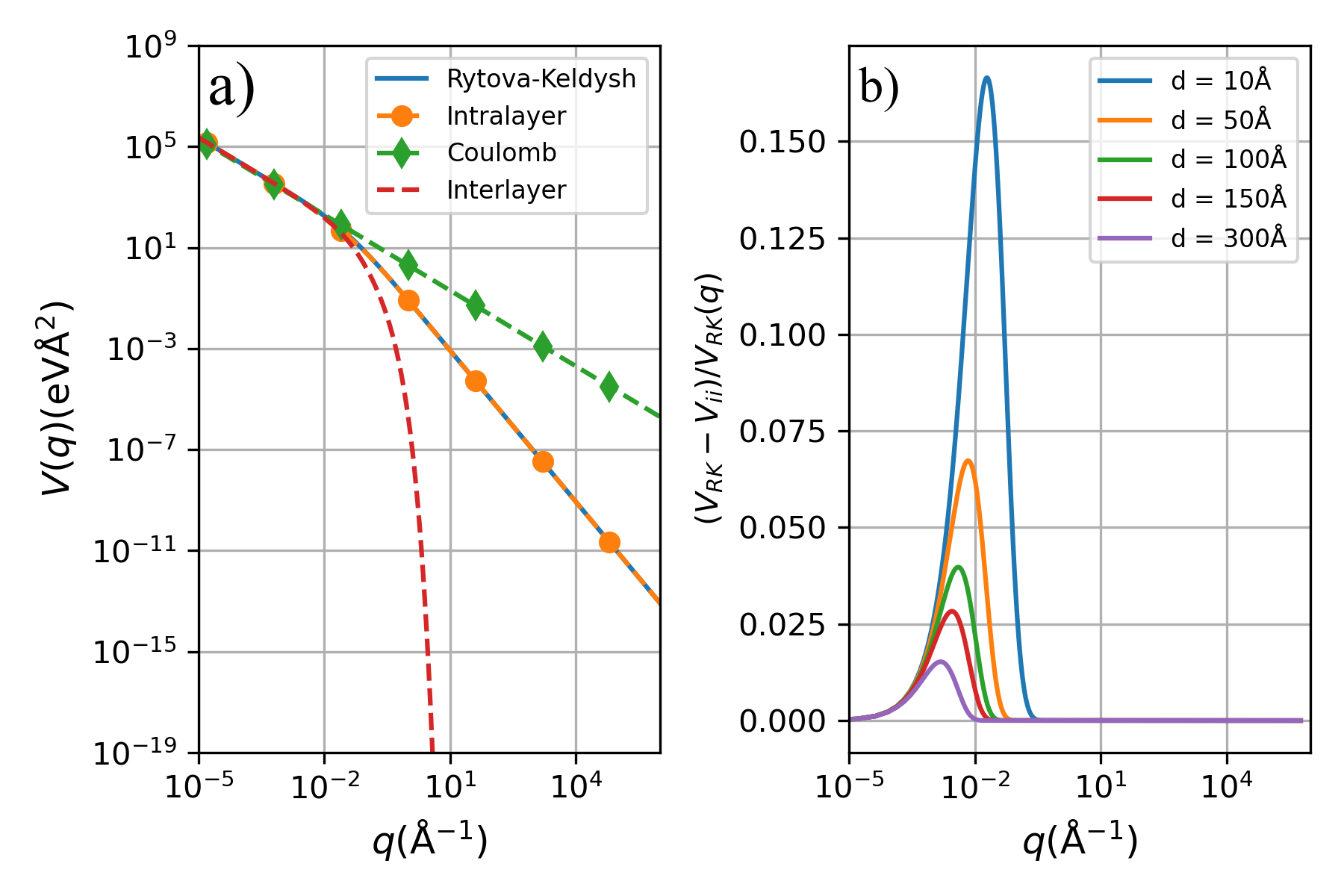}
\caption{(Color online) (a) Comparison between different carrier-carrier potentials in momentum space: RK (solid blue curve), Coulomb (dashed green curve with rhombus symbols), intralayer (dashed orange curve with circular symbols), and interlayer (dashed red curve) interactions. The interlayer [Eq.~\eqref{eq_Viipot}] and the intralayer [Eq.~\eqref{eq_Vijpot}] potentials were calculated considering $r_1=r_2=r=44.68 $~\AA\ and  $d=7.15 $~\AA. When $q$ is of the order of $1/r$, the Coulomb potential deviates from the other three and a negligible difference between the intralayer and the RK potentials is observed. The interlayer potential shows a strong screening that is due to the term proportional to $e^{-qd}$ of $G_j(q)$ in Eq.~\eqref{eq_Gq} when $q\approx 1/d$. (b) The relative difference between the intralayer and the RK potentials, which can be as high as 15$\%$ the shorter the spacer width $d$.}
\label{Fig2}
\end{figure}

The carrier-carrier interaction was derived from the Poisson equation in Appendix \ref{subsec.poisson} with the geometry presented in Fig.~\ref{Fig1}(a) for both intralayer $V_{ii}$ and interlayer $V_{ij\neq i}$ potentials, defined as the interaction between carriers in the same (intra) layer or in adjacent (inter) layers, and respectively given by [\onlinecite{ruiz2020theory}]
\begin{subequations}
\begin{eqnarray}
V_{ii}(q) &=& \frac{-e^2}{q\epsilon_{0}\left[\epsilon_{1}+r_{i}q+\epsilon_{2}G_j(q)\right]},
\label{eq_Viipot} \\
V_{ij\neq i}(q) &=& V_{ii}(q)\left[\cosh(qd)-G_j(q)\sinh(qd)\right], \label{eq_Vijpot}
\end{eqnarray}
\end{subequations}
where
\begin{equation}
G_j(q)=\frac{\cosh(qd)(\epsilon_{3}+r_{j}q)+\epsilon_{2}\sinh(qd)}{\epsilon_{2}\cosh(qd)+\sinh(qd)(\epsilon_{3}+r_{j}q)}, \label{eq_Gq}
\end{equation}
with $r_i$ being the screening length of each 2D layer and $i=\{1,2\}$. Fig.~\ref{Fig2}(a) shows a comparison between different interactions in momentum space: Rytova-Keldysh [RK - Eq.~\eqref{eq_RK}], Coulomb, interlayer [$V_{i\neq j}$ - Eq.~\eqref{eq_Viipot}] and intralayer [$V_{ii}$ - Eq.~\eqref{eq_Vijpot}] potentials. Although, they converge to the same value in the long-wavelength limit, \textit{i.e.} when $q\approx 1/d$, the interlayer potential deviates from the RK and intralayer potentials. Fig.~\ref{Fig2}(b) emphasizes the difference between the intralayer and RK interactions magnitudes, showing a difference of almost $15\%$ between them for a short spacer width.

\subsection{Chebyshev method}\label{subsec.tcheb}

The carrier-carrier interaction in the classical regime will diverge in the infrared limit, which must be handled to solve the Wannier equation numerically in momentum space. Here, we use the method developed by Chawla and Kumar \cite{golberg2013numerical} to analytically remove this infrared divergence of the kernel by expanding in Chebyshev polynomials and analytically integrating out the divergence via Cauchy principal value. 

We start with the Wannier equation in momentum space:
\be
E_p\psi(\bp)+\int \frac{d\bp'}{(2\pi)^2}
V(\bp-\bp')\psi(\bp')=E\psi(\bp), \label{eq_sch_partial}
\ee
that also corresponds to a simplified version of the Bethe-Salpeter equation in the ladder approximation, when neglecting the exchange term for a two-band system in the effective mass regime. Decomposing Eq.~\eqref{eq_sch_partial} in partial waves, we have
\be\label{eq_sch_partial_new}
E_p\psi_\ell({p})+ \frac{1}{2\pi} \int_0^\infty dp'  p'
V_\ell(p,p')\psi_\ell(p')=E\psi_\ell(p),
\ee
with the interaction given by
\be
V_\ell(p,p')= \frac{1}{2\pi} \int_0^{2\pi} d\phi V(p-p',\phi) \cos(\ell\phi).
\ee
Now, we consider the hyperbolic conformal mapping
\be \label{hypmap}
u=\frac{\xi p-1}{\xi p+1},
\ee
with $u\in [-1,1]$ and $\xi$ being a scale parameter, and expand the momentum space wave function in Chebyshev polynomials $T_n$, such as
\be
\psi_\ell(u)=f(u)\sum_n c_{n,\ell} T_n(u), \label{eq_tcheb}
\ee 
where $f(u)$ is a function used to speed up the convergence. The choice of $f(u)$ shall be discussed later on. Writing the integrand of Eq.~\eqref{eq_sch_partial_new} in terms of $u$, one has
\begin{multline}
 \frac{1}{2\pi} \int_0^\infty p' dp' V_\ell(p,p')\psi_\ell(p')
=\\
 \frac{1}{\xi^2}\int_{-1}^{1}du^\prime \frac{V_\ell(u,u^\prime)\psi_\ell(u^\prime)(1+u^\prime)}{\pi(1-u^\prime)^3}. \label{eq_integral}
\end{multline}

From the electrostatic nature of the RK potential, one of the numerically slow-step in solving Eq.~\eqref{eq_sch_partial_new} comes from the $1/q$ infrared singularity, that we shall demonstrate how it can be analytically removed. Now, introducing the expansion given  by Eq.~\eqref{eq_tcheb} in Eq.~\eqref{eq_integral}, one obtains
\be
I_{n,\ell}(u)=\frac{1}{\xi^2}\int_{-1}^1du^\prime \frac{V_\ell(u,u^\prime)(1+u^\prime)}{\pi(1-u^\prime)^3}f(u^\prime)T_n(u^\prime), 
\ee
where now the $1/q$ infrared singularity appears explicitly when $u=u^\prime$:
\be
I_{n,\ell}(u)=\frac{1}{\xi^2}\int_{-1}^1 du^\prime \frac{K_\ell(u,u^\prime) T_n(u^\prime)}{u-u^\prime}, \label{eq_In} 
\ee
with the kernel being set to
\begin{eqnarray}
K_\ell(u,u^\prime)=\frac{V_\ell(u,u^\prime)(1+u^\prime)}{\pi(1-u^\prime)^3}f(u^\prime)(u-u^\prime), \label{eq_kernel}
\end{eqnarray}
which vanishes for $u=u^\prime$. By a careful analysis of Eq.~\eqref{eq_kernel}, one has that a convenient choice for the function $f(u)$ is
\be
f(u)=\frac{1-u^3}{1+u},
\ee
which removes the pole at $u=1$ in the kernel and will be used to compute the exciton eigenstates in Sec.~\ref{sec_results}.

Now, we use Chawla and Kumar's method \cite{golberg2013numerical} to compute the integral in Eq.~\eqref{eq_In}. Decomposing the kernel, Eq.~\eqref{eq_kernel}, in Chebyshev polynomials, one gets
\be
K_\ell(u,u^\prime)\approx \sum_{j=0}^M b_j(u)T_j(u^\prime),
\ee
where $b_j(u)$'s are the expansion coefficients. Analytically integrating Eq.~\eqref{eq_In}, one obtains
\be
I_{n,\ell}(u)=\frac{1}{2\xi^2}\sum_{j=0}^M b_j(u)\left[\lambda_{j+n}(u)+\lambda_{|j-n|}(u) \right],
\ee
where the $\lambda_k(u)$ function is defined in the Appendix \ref{app_CK} and can be obtained recursively. Replacing back in Eq.~\eqref{eq_integral}, we have that
\begin{equation}
\sum_{n=0}^\infty\left[h(u)f(u) T_n(u)+ I_{n,\ell} -Ef(u) T_n(u) \right]c_{n,\ell}=0, \label{eq_final}
\end{equation}
where 
\be 
h(u)=\frac{\hbar^2}{2\mu\xi^2}\left(\frac{1-u}{1+u} \right)^2\, .\ee
Truncating the expansion at a maximum value $n=N$, we can solve Eq.~\eqref{eq_final} as a linear homogeneous system (generalized eigenvalue problem) by choosing $N+1$ different values for $u$. For this, we can choose the zeros of the $T_{N+1}$ Chebyshev polynomial.

\subsection{Bandgap engineering}\label{sec.bandgap}

The quasiparticle band structure of 2D materials depends on the dielectric environment~\cite{chaves2020bandgap}. To account for this dependence, we employ the Semiconductor Bloch Equations (SBE)\cite{kira2011semiconductor} for the heterostructure depicted in Fig.~\ref{Fig1}(a). We neglect the tunneling between the MX$_2$ layers due to the presence of a dielectric spacer between them. The single particle Hamiltonian for the charge carriers in each layer can be described by the following massive Dirac equation\cite{Kormanyos_2015}
\be
\hat{H}_{0,i}=\tau_i\hbar v_{F,i} \boldsymbol{\sigma}\cdot\mathbf{p}_i+\sigma_{z}\Delta^0_{s,\tau},
\ee
whose the mass term $\Delta^0_{s,\tau}$, corresponding to the bare ``bandgap'', depends on the spin ($s$) and valley ($\tau$) indexes for each layer $i$. $v_{F,i}$ denote the Fermi velocity of the layer $i$ and $\sigma_z$ is the $z$ Pauli matrix component.

\begin{figure}[t]
\centering
\includegraphics[width=0.5\columnwidth]{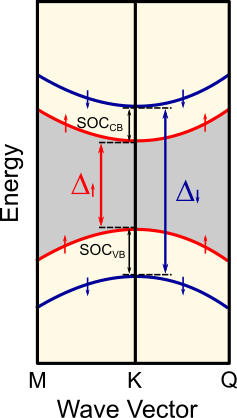}
\caption{(Color online) Schematic illustration of the lowest conduction (CB) and valence (VB) bands of monolayer TMDs in the vicinity of the $K$ (red curves) and $K^\prime$ (blue curves) points, emphasizing the band splitting due to SOC and spin flipping for each band in the opposite valley due to the inversion symmetry. The up (red) and down (blue) arrows stand for spin-up and spin-down states. SOC$_{CB}$ (SOC$_\mathrm{VB}$) corresponds to the energetic split of the conduction (valence) band.}
\label{fig_band_schm}
\end{figure}

In order to take into account the corrections to the bandgap, we employed the procedure derived in Ref.~[\onlinecite{Chaves_2017}], by considering the aforementioned gapped Dirac equation, the electron-electron interaction, and a dipole coupling with light. It is well-known that TMDs have a strong spin-orbit coupling (SOC) originating from the $d$ orbitals of the metal atoms and, consequently, it induces a spin splitting of bands in monolayer,\cite{liu2015electronic} as illustrated in Fig.~\ref{fig_band_schm}. Thus, by applying Heisenberg's equation to the polarization operator, we arrive at the following exchange self-energy expression for each layer $j$, given by the Random Phase-Approximation (RPA) for 2D massive Dirac Hamiltonian \cite{Chaves_2017}, such as
\be
\Sigma^j_{s\tau}(\bk)=\int \frac{d\mathbf{q}}{4\pi^2} V_{jj}(q) n_{s\tau}(\bk-\mathbf{q})\frac{4\hbar^2v_F^2\bk\cdot\mathbf{q}+(\Delta^0_{s,\tau})^2 }{4E^{s\tau}_{jk} E^{s\tau}_{jq}}, \label{eq_self_energy}
\ee
from which we can calculate the dressed bandgap as 
\be
\Delta^j_{s\tau}=\Delta^j_{s\tau,0}+\Sigma_{s\tau}^j(k=0),\label{eq_dressed}
\ee
where $\Delta^j_{s,\tau}$ denotes the energy difference between the conduction and valence bands with the same $s$ and $\tau$ indexes for each layer $j$ at the $K$ point, the intralayer potential $V_{jj}$ is given by Eq.~\eqref{eq_Viipot}, $n_{s\tau}$ is the valence electronic density, and $E_{jk}^{s\tau}$ is the eigenvalue of the massive 2D Dirac Hamiltonian. The intralayer interaction depends on the dielectric environment through the spacer width $d$, the dielectric constants $\epsilon_{i}$, and the monolayer screening lengths $r_i$. As our goal is to study the dependence of the exciton properties on the system geometry, we fit the monolayers screening length $r_0$ to reproduce the experimental exciton energy of the suspended monolayer for each MX$_2$ as described in Appendix~\ref{app_fitt}. 

\begin{figure}[t]
  \centering
  \includegraphics[width=1.0\linewidth]{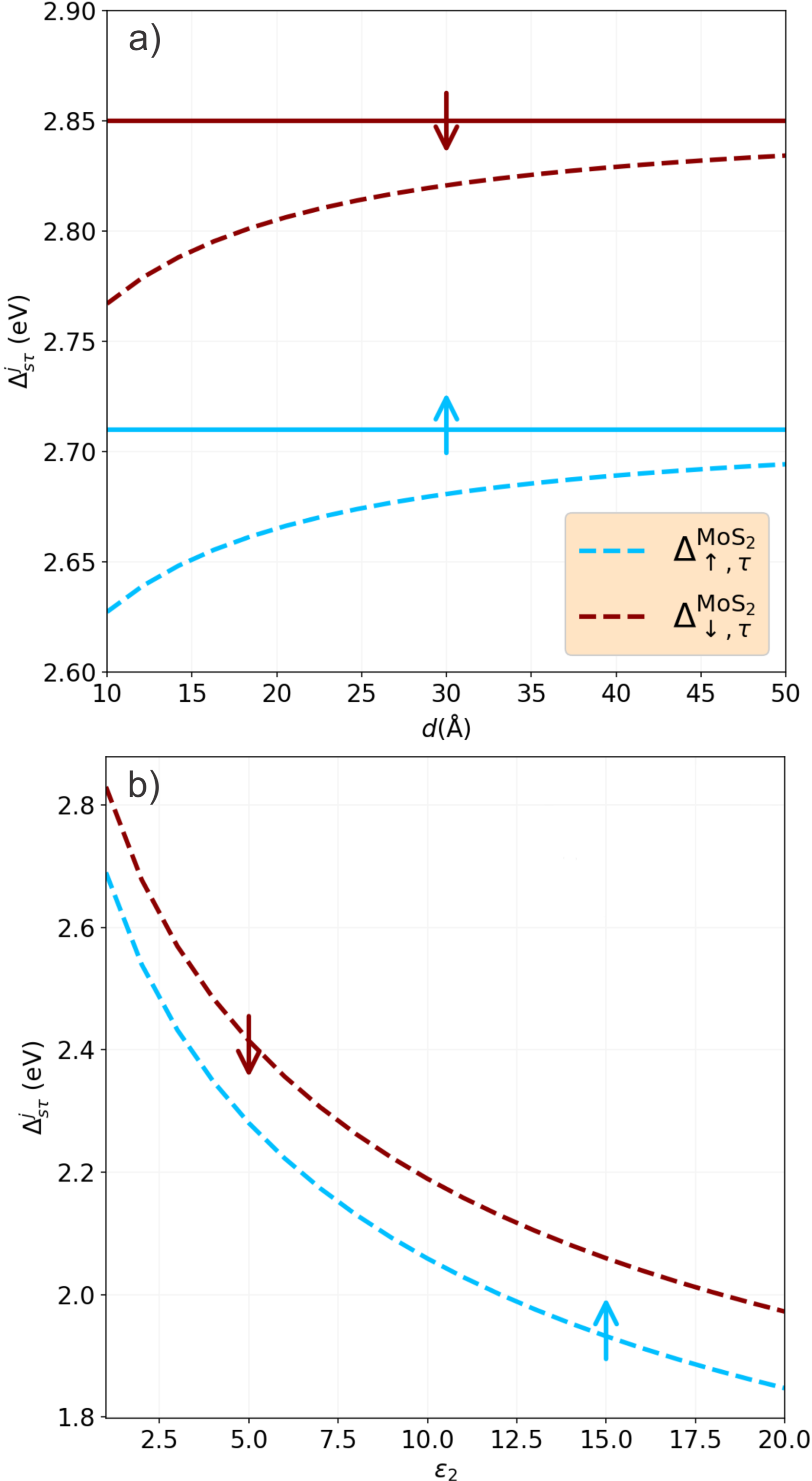}
  \caption{(Color online) $K-K$ transition energies of both spins for MoS$_{2}$ at the MoS$_2$/MoSe$_{2}$ heterostructure with respect to the changes (a) in the interlayer separation $d$ in a suspended sample with $\epsilon_1=\epsilon_2=\epsilon_3=1$, and (b) in the spacer dielectric constant $\epsilon_{2}$ with a fixed interlayer distance $d = 7.15$ \AA\ and external dielectric constants $\epsilon_1= \epsilon_3 = 1$. Cyan and red curves correspond to up $\left(\Delta^{MoS_2}_{\uparrow,\tau}\right)$ and down $\left(\Delta^{MoS_2}_{\downarrow,\tau}\right)$ spin results, respectively. The solid lines in (a) represent a monolayer limit ($d \to \infty$) of the MoS$_2$.}
  \label{fig_gapxd}
\end{figure}

Using the $r_0$'s given in Table~\ref{tab_fit} in Appendix~\ref{app_fitt}, we obtain the bare bandgap $\Delta^j_{s\tau,0}$ from Eq.~\eqref{eq_dressed} for each material and spin-valley combination for the suspended monolayer, \textit{i.e.} for $\epsilon_1=\epsilon_2=\epsilon_m=1$ and $d\rightarrow \infty$. The obtained values are presented in Table~\ref{tab_gaps}. With the fitted values of $r_0$ and $\Delta_{s\tau}^0$, we can solve Eq.~\eqref{eq_self_energy} for different geometric setups and study the dependence of the $\Delta_{s\tau}^j$, \textit{i.e.} the spin/valley dependent transition energy at the $K$ point. In Fig.~\ref{fig_gapxd}(a), we show that the mutual electrostatic screening between two monolayers can decrease the value of $\Delta^j_{s\tau}$ by $50$~meV as the interlayer separation decreases to $7.15$\,\AA. In Fig.~\ref{fig_gapxd}(b), we show the dependence of $\Delta^j_{s\tau}$ on the spacer dielectric constant. The huge renormalization of the bandgap due to the electron-electron interaction~\cite{ugeda2014giant} is weakened by the spacer dielectric screening, and as the dielectric constant is increased, the transition energy approaches the bare value $\Delta_{s\tau,0}$. In Figs.~\ref{fig_gapxd}(a) and \ref{fig_gapxd}(b) it was assumed the MoS$_2$/MoSe$_2$ heterostructure, however qualitatively similar results are expected for the other different TMD layer compound combinations.

\begin{table}[!h]
\caption{\textit{Ab initio} bandgaps \cite{Zhang_2016}, Fermi velocity \cite{Kormanyos_2015} and calculated bare bandgaps using Eq.~\eqref{eq_dressed} and the fit $r_0$'s given in Table~\ref{tab_fit} in Appendix~\ref{app_fitt} for the four investigated TMDs and different combinations of spin and valley indexes.} \label{tab_gaps}
\begin{center}
 \begin{tabular}{l| c c c c c} \hline \hline
 Materials & $\Delta_\uparrow$ (eV) & $\Delta_\downarrow$ (eV)& $v_F$ (eV $\cdot$ \AA)& $\Delta_\uparrow^0$(eV) & $\Delta^0_\downarrow$(eV) \\ \hline
 MoS$_2$ & 2.71 & 2.85 & 2.76 & 1.29 & 1.39 \\ 
 MoSe$_2$ & 2.37  & 2.55 & 2.53 & 1.18 & 1.32  \\
 WS$_2$& 2.96 & 3.30 & 3.34 & 1.35 & 1.61\\
 WSe$_2$ & 2.63 & 3.01 & 3.17 & 1.14 & 1.40\\ \hline\hline
\end{tabular}
\end{center}
\end{table}

\section{Results}\label{sec_results}

\begin{figure}[t]
\centering
\includegraphics[width=1.0\columnwidth]{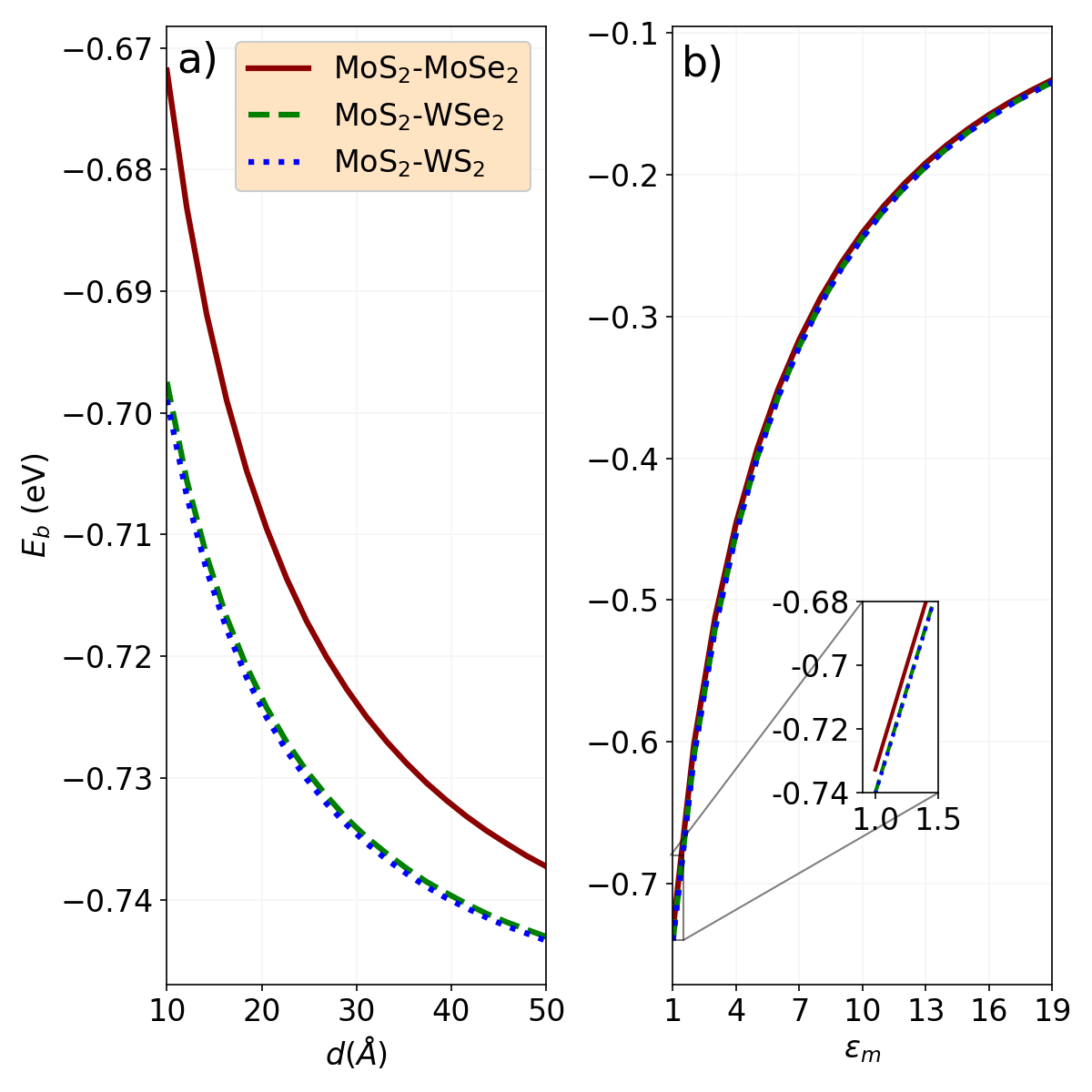}
\caption{(Color online) Binding energies ($E_B$) of the intralayer A excitons, referred to as an electron-hole pair lying in the MoS$_2$ layer, by taking different layer compounds in the TMD heterostructure formation. Red solid, green dashed, and blue dotted curves correspond to $\mbox{MoS}_2 - \mbox{MoSe}_2$, $\mbox{MoS}_2 - \mbox{WSe}_2$, and $\mbox{MoS}_2 - \mbox{WS}_2$ double-layers, respectively. Panels (a) and (b) show the dependence of $E_B$ on the separation distance of the layers $d$, by assuming $\epsilon_1=\epsilon_2=\epsilon_m=1$, and on the dielectric constant $\epsilon_m$, by assuming a fixed interlayer distance of $d=41$~\AA\ and dielectric constants of the substrate and superstrate as $\epsilon_1=\epsilon_2=1$, respectively. An enlargement as an inset in panel (b) emphasizes the small energetic difference between the binding energies for the $\mbox{MoS}_2 - \mbox{MoSe}_2$ heterojunction and the other two, $\mbox{MoS}_2 - \mbox{WSe}_2$ and $\mbox{MoS}_2 - \mbox{WS}_2$, double-layers.}
\label{fig_mos2_heteros}
\end{figure}

Based on the formalism presented in the previous sections, in the current section, we shall discuss the exciton wave functions and energies, as well as the binding energies, for different combinations of double-layer TMD heterostructures. For that, we solve the truncated Eq.~\eqref{eq_final} using the carrier-carrier potentials given by Eq.~\eqref{eq_Viipot} for the case of intralayer excitons and by Eq.~\eqref{eq_Vijpot} for the interlayer excitons. All system parameters assumed here for each one of the four investigated TMDs that composes the double-layer are expressed in Tables~\ref{tab_gaps} and \ref{tab_fit}, as for instance the effective masses, material's bandgap, and the 2D material screening length $r_0$ that was fitted to give the exciton binding energy as explained in Appendix~\ref{app_fitt}. It is worth mentioning that tunneling effects of the charge carriers between the two layers are neglected here, \textit{i.e.} we consider the approximation that the electron and hole wave functions of each TMD layer do not overlap.

\begin{figure}[t]
\centering
\includegraphics[width=1.0\columnwidth]{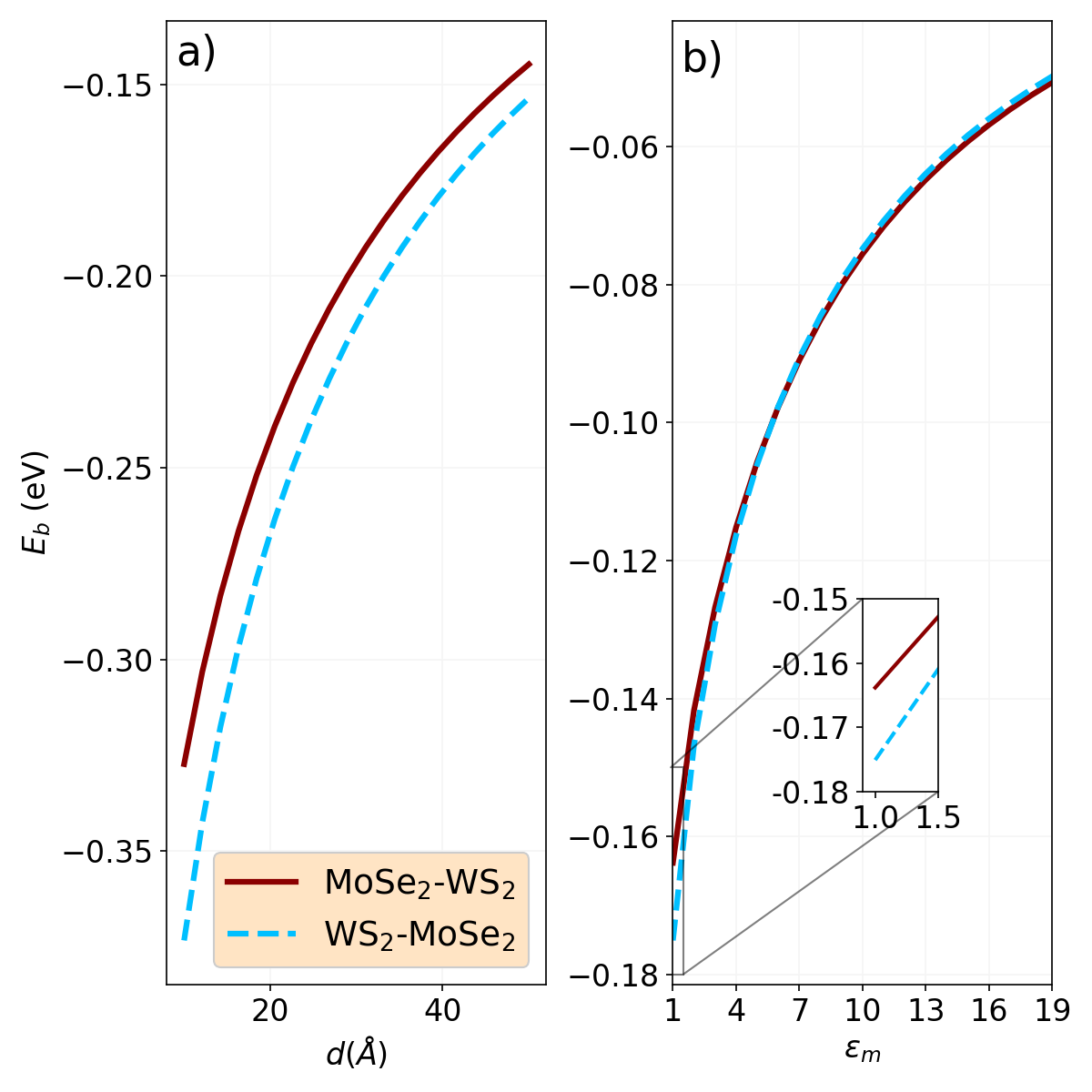}
\caption{(Color online) Binding energies ($E_B$) of the interlayer excitons in the MoSe$_2$ layer by taking different layer compounds in the TMD heterostructure formation. Red solid and cyan dashed curves correspond to $\mbox{MoSe}_2 - \mbox{WS}_2$ and $\mbox{WS}_2 - \mbox{MoSe}_2$, respectively, with the interlayer exciton being formed by the electron (hole) of the first (second) referred compound. Panels (a) and (b) show the dependence of $E_B$ on the separation distance of the layers $d$, by assuming $\epsilon_1=\epsilon_2=\epsilon_m=1$, and on the dielectric constant $\epsilon_m$, by assuming a fixed interlayer distance of $d=41$~\AA\ and dielectric constants of the substrate and superstrate as $\epsilon_1=\epsilon_2=1$, respectively. An enlargement as an inset in panel (b) emphasizes the energetic difference between the binding energies for the $\mbox{MoSe}_2 - \mbox{WS}_2$ and $\mbox{WS}_2 - \mbox{MoSe}_2$ double-layers.}
\label{fig_inter_heteros}
\end{figure}

Figures~\ref{fig_mos2_heteros}(a) and \ref{fig_mos2_heteros}(b) show the binding energy of the intralayer A excitons, which are formed when the electron-hole pair lies on the MoS$_2$ layer, as a function of the separation distance (spacer width) $d$ and the dielectric constant of the spacer $\epsilon_m$, respectively. Results for three different layer compounds in the heterostructure formation are shown: (red solid curve) $\mbox{MoS}_2 - \mbox{MoSe}_2$, (green dashed curve) $\mbox{MoS}_2 - \mbox{WSe}_2$, and (blue dotted curve) $\mbox{MoS}_2 - \mbox{WS}_2$. As a consequence of the fact that MoSe$_2$ has the larger $r_0$ value (see Table~\ref{tab_fit}) of the four investigated TMD layers, it was already expected that it would screen more effectively the electron-hole interaction by the charge-image effect. As verified in Fig.~\ref{fig_mos2_heteros}(a), it lowers the exciton binding energy by almost 20\,meV, whereas the WSe$_2$ and WS$_2$ cases present almost identical binding energies due to their very similar $r_0$ values. From Fig.~\ref{fig_mos2_heteros}(b), one notices that the intralayer A exciton binding energies are strongly affected by the spacer's dielectric constant $\epsilon_m$ changes, exhibiting an energetic variation on the order of 300\,meV when $\epsilon_m$ varies from $1$ to $4$. Qualitatively similar results were reported in the TMD monolayer case in Refs.~[\onlinecite{ilkka2015,chaves2020bandgap}], being physically understood by the spatial localization of the interlayer A exciton depicted in Fig.~\ref{fig_mos2_heteros} that lies only in one of the layers of the double-layer TMD system. Moreover, a small energetic difference of the order of a few meV is noted in Fig.~\ref{fig_mos2_heteros}(b) for the binding energies of the intralayer A excitons in the MoS$_2$ when one compares the different investigated heterostructures. It is emphasized by the enlargement shown as an inset of Fig.~\ref{fig_mos2_heteros}(b). It reveals structural independence in the heterostructure formation on the binding energy as a function of the dielectric constant, \textit{i.e.} $\epsilon_m$ changes similarly affect the binding energies regardless of the adjacent TMD layer of the MoS$_2$-formed heterostructure.

\begin{figure}[t]
\centering
\includegraphics[width=1.0\columnwidth]{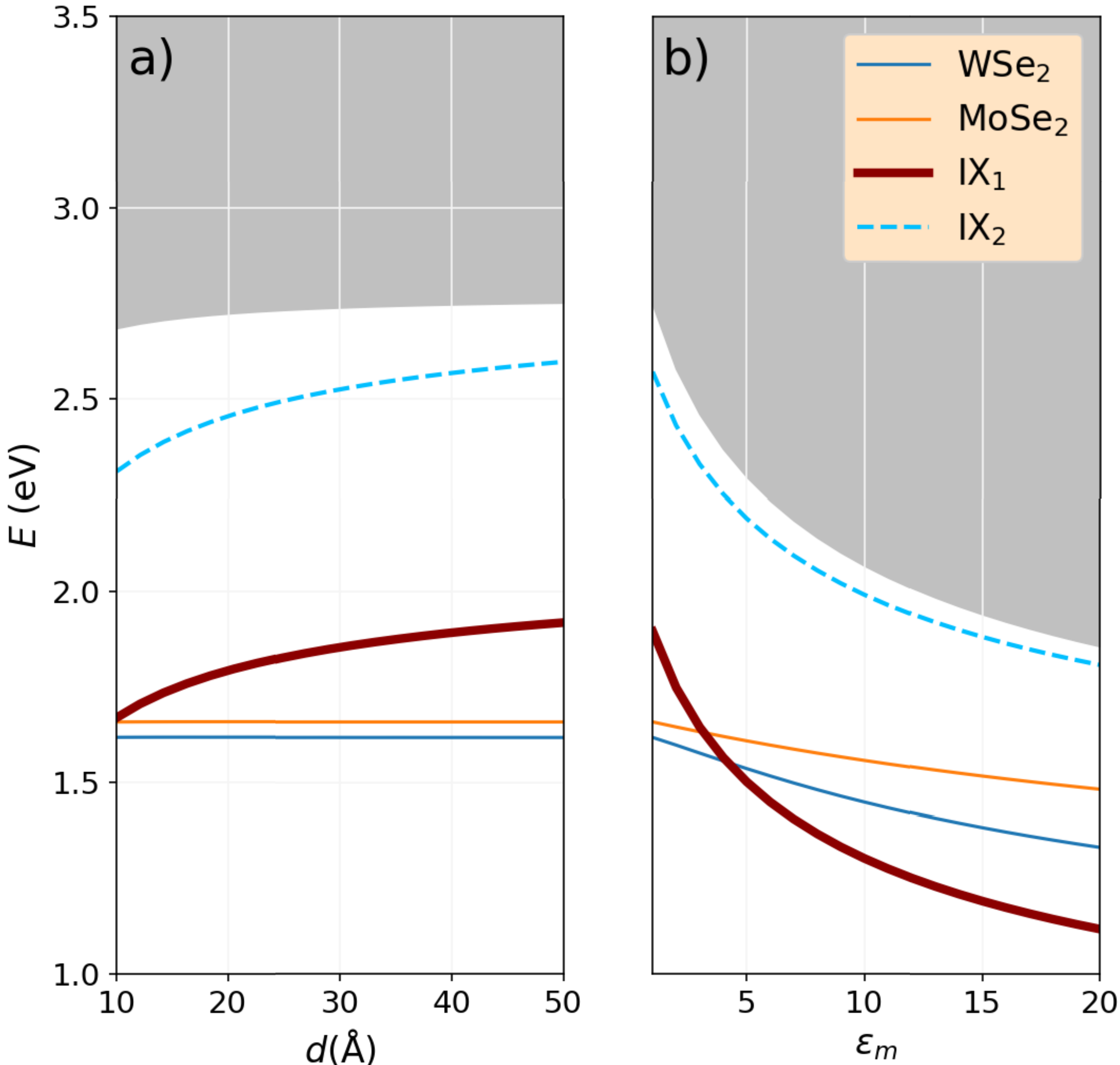}
\caption{(Color online) Exciton energy dependency on (a) the layer separation and (b) the dielectric media $\epsilon_{m}$ for the MoSe$_2$-WSe$_2$ heterostructure. IX$_i$ denotes the $i$-th interlayer exciton, such that IX$_1$ (IX$_2$) is formed by the electron from the lowest conduction band of the first (second) material and the hole from the highest valence band of the second (first) material with result represented by the solid red (dashed cyan) curve. Solid blue and yellow curves correspond to the intralayer excitons for WSe$_2$ and MoSe$_2$ cases, respectively. (a) All dielectric constants are held fixed with the value of $1$, and (b) the layer separation is fixed to $d=41\,$\AA. The shaded gray region corresponds to the continuum.}
\label{fig_mose2wse2}
\end{figure}

Let us now focus on the interlayer exciton. When stacking different TMD monolayers, the corresponding Dirac $K$ points in the reciprocal space of each TMD monolayer will not coincide, and the distance between the respective $K$ points of each layer depends both on the relative rotation of the crystallography orientation and the mismatch of the lattice parameters of each layer. Here, within the effective mass approximation, we are ignoring both effects. Considering only the uppermost valence band and the lowest conduction band of each layer, there are two different kinds of interlayer excitons for the type II band alignment case (see Fig.~\ref{Fig1}): (i) the lowest conduction band between the two 2D materials hosting the electron, whereas the hole is hosted in the valence band of the adjacent layer that possesses the highest energy, and (ii) the opposite formation, \textit{i.e} the highest conduction band between the TMD monolayers hosting the electron, whereas the hole is hosted in the valence band of the adjacent layer that possesses the lowest energy. If the corresponding exciton binding energy has a magnitude smaller than the conduction band offset, this will result in an excitonic resonance, as the exciton energy lies inside the conduction band.

Results for these two mentioned kinds of interlayer excitons in double-layer heterostructures composed by $\mbox{MoSe}_2$ and $\mbox{WS}_2$ compounds are shown in Fig.~\ref{fig_inter_heteros}. The solid red (dashed cyan) curve corresponds to the interlayer exciton formed by an electron (hole) from the MoSe$_2$ (WS$_2$) and a hole from the WS$_2$ (MoSe$_2$). Both interlayer exciton configurations show a binding energy increase when the layer separation $d$ decreases, attaining values of almost $400$~meV for shorter distances of the order of $10$~\AA\ [see Fig.~\ref{fig_inter_heteros}(a)]. Such behavior is easily understood by the electrostatic interaction nature of the electron-hole attraction, which is enhanced the shorter the interlayer distance. One also observes in Fig.~\ref{fig_inter_heteros}(a) that the energetic difference of the binding energies for the two configurations of interlayer excitons, \textit{i.e.} $|E_b^{\mbox{MoSe}_2-\mbox{WS}_2} - E_b^{\mbox{WS}_2-\mbox{MoSe}_2}|$, increases when the interlayer distance decreases. Knowing that the interlayer interaction depends on the layer separation and the screening parameters $r_0$ of heterostructures' compounds, and in addition to that, here we are switching the layers where the electron and hole are positioned, one can link this energetic difference $|E_b^{\mbox{MoSe}_2-\mbox{WS}_2} - E_b^{\mbox{WS}_2-\mbox{MoSe}_2}|$ in view of the interlayer exciton formation and the consequent overall strength switching of the role of the electrostatic interaction at each layer. Note that the electrostatic interaction of an electron-hole pair separated by a dielectric media has its amplitude modulated by the electrostatic screening of the layers damped by the separation between them. Thus, by exchanging the configuration of the electron-hole layer location, one leads to dampening/enhancing the screening of the adjacent layer owing to the layer separation and consequently to an energetic difference in the binding energy of the exciton. A similar feature is observed in the case that we fixed the layer separation and vary the dielectric constants of the environment. This is present in Fig.~\ref{fig_inter_heteros}(b). Note that the interlayer exciton binding energy exhibits the same tendency as the intralayer one [see Fig.~\ref{fig_mos2_heteros}(b)] as a function of the spacer dielectric constant $\epsilon_m$, except for the increased energetic distancing between the two $\mbox{MoSe}_2-\mbox{WS}_2$ and $\mbox{WS}_2-\mbox{MoSe}_2$ cases when $\epsilon_m$ assumes high values, as emphasized in the inset of Fig.~\ref{fig_inter_heteros}(b).

In what follows, we study the exciton energy, which is defined by
\begin{equation}
    E_\mathrm{exc}=E_c-E_v-|E_b|, 
\end{equation}
where $|E_b|$ is the magnitude of the exciton binding energy, $E_c$ the bottom of the conduction band, and $E_v$ the top of the valence band associated with the electron and hole, respectively, that contributes to the exciton formation. For a bright exciton, this value also corresponds to the energy of the photon that creates the electron-hole bound-state. 

\begin{figure}[t]
\centering
\includegraphics[width=1.0\columnwidth]{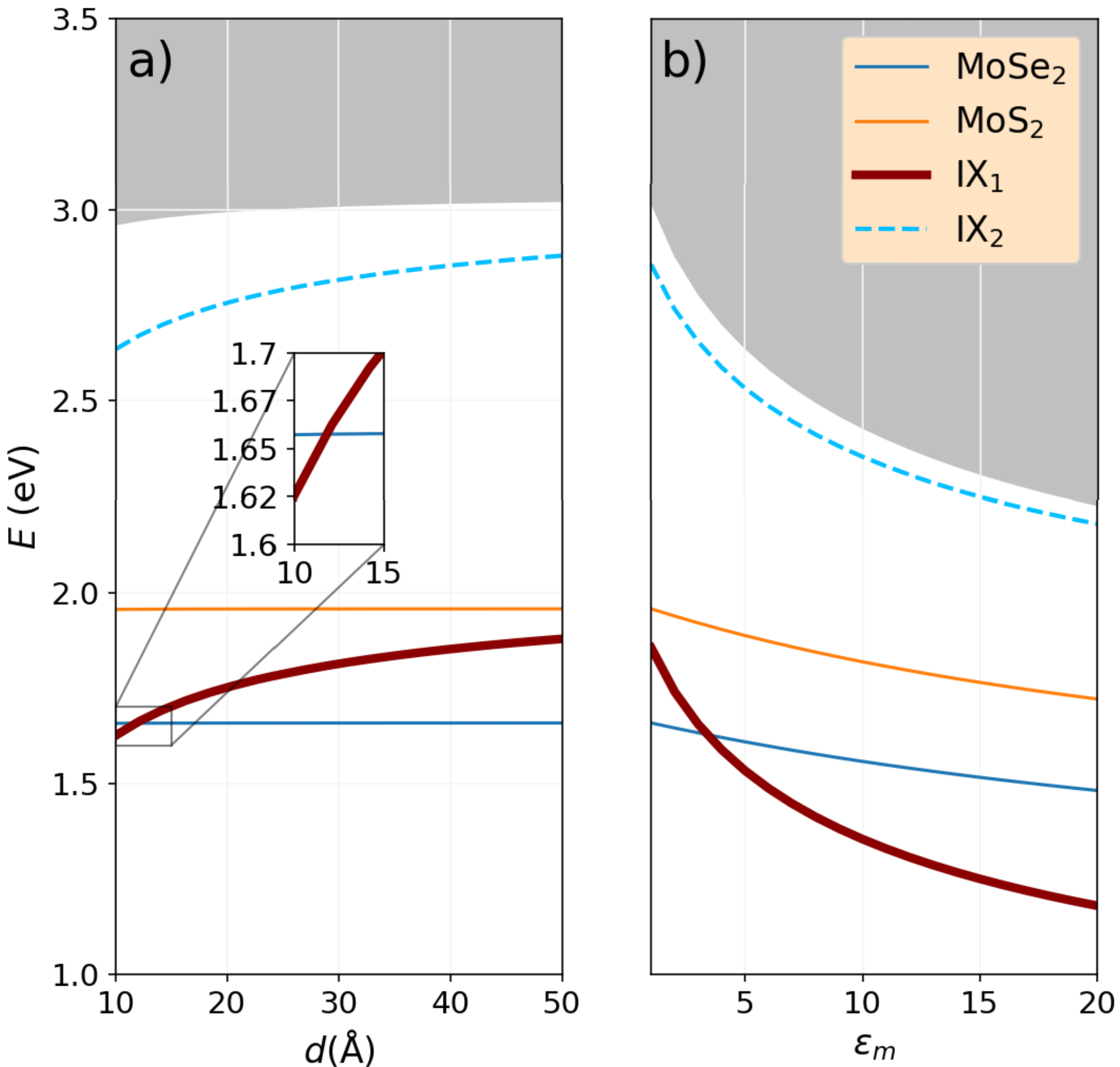}
\caption{(Color online) Exciton energy dependency on (a) the layer separation and (b) the dielectric media $\epsilon_{m}$ for the MoS$_2$-MoSe$_2$ heterostructure. IX$_i$ denotes the $i$-th interlayer exciton, such that IX$_1$ (IX$_2$) is formed by the electron from the lowest conduction band of the first (second) material and the hole from the highest valence band of the second (first) material with result represented by the solid red (dashed cyan) curve. Solid blue and yellow curves correspond to the intralayer excitons for MoSe$_2$ and MoS$_2$ cases, respectively. (a) All dielectric constants are held fixed with the value of $1$, and (b) the layer separation is fixed to $d=41\,$\AA. The shaded gray region corresponds to the continuum. An enlargement around small layer separation is shown as an inset of panel (a).}
\label{fig_MoS2mose2}
\end{figure}

\begin{figure}[t]
\centering
\includegraphics[width=1.0\columnwidth]{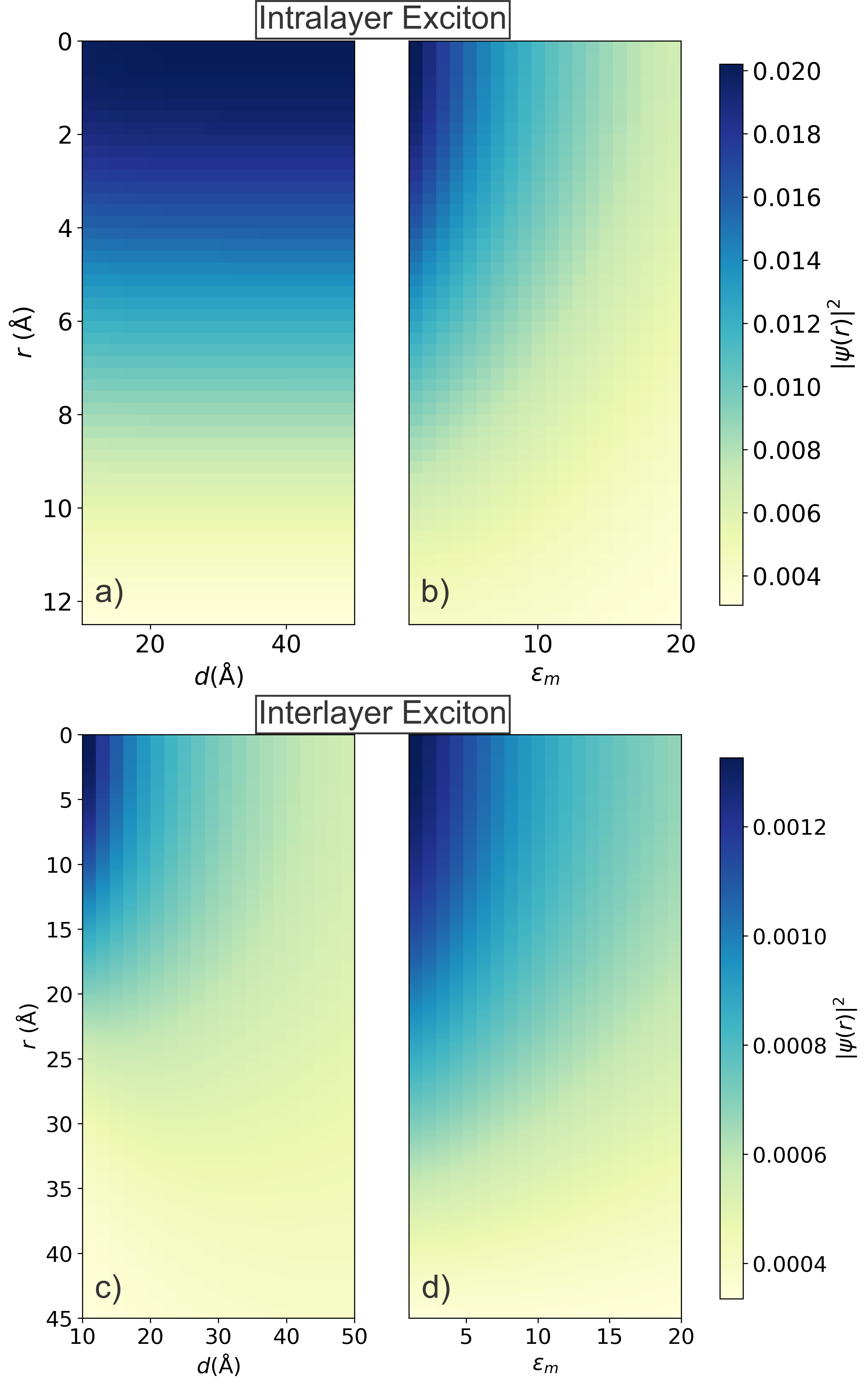}
\caption{(Color online) (a,b) Intralayer and (c,d) interlayer exciton wave function for the MoS$_2$-MoSe$_2$ heterostructure as a function of (a,c) the layer separation and (b,d) the dielectric constant. The dielectric constants are held fixed at $1$ for panels (a) and (c), whereas the value for the interlayer distance is fixed of $d=41$ \AA\ in panels (b) and (d).}
\label{fig_IntralayerMap_InterlayerMap}
\end{figure}

From now on, for an MX$_{2}$-M$^{\prime}$X$^{\prime}_{2}$ heterostructure, we define the  interlayer exciton IX$_1$ as the bound-state of the electron from the lowest conduction band of the first material and the hole from the highest valence band of the second material and IX$_2$, as the opposite. In Fig.~\ref{fig_mose2wse2}, we show the evolution of the exciton energies, both intralayer and interlayer, and the bottom value of the conduction band as a function of [Fig.~\ref{fig_mose2wse2}(a)] the interlayer spacing and [Fig.~\ref{fig_mose2wse2}(b)] the dielectric constant of the spacer. It is worth mentioning that we use as a reference energy level the top of the valence band, considering the band alignment of Ref.~[\onlinecite{Zhang_2016}]. One can see in Fig.~\ref{fig_mose2wse2}(a) that the intralayer exciton energies (solid blue and yellow curves for WSe$_2$ and MoSe$_2$, respectively) are very robust with respect to the layer separation due to the simultaneous changes of the bandgap and the exciton binding energy, which cancel each other out, keeping the energies of the intralayer exciton unaltered. As the interlayer separation $d$ increases, the value of each intralayer exciton energy converges to the suspended monolayer value minus the band alignment energy. For the interlayer exciton (see solid red and dashed cyan curves for IX$_1$ and IX$_2$, respectively), we have that the exciton energy increases due to the weakening of the binding energy, which arises from the sensitivity of the interlayer interaction with respect to the layer separation. For instance, notice in Fig.~\ref{fig_mose2wse2}(a) that the interlayer exciton IX$_1$ energy (dashed cyan curve) increases $0.15\,$eV for $d=50$ \AA. By Fig.~\ref{fig_mose2wse2}(b), one observes that the intralayer exciton energy is more sensitive to changes in the dielectric media. By increasing the dielectric constant of the space $\epsilon_m$, the screening is enhanced and, therefore, weakening the Coulomb interaction. Although the interlayer exciton binding energy varies less with respect to the dielectric screening, the gap correction is more acute, leading to a larger fluctuation of the interlayer exciton energy.

Similarly to Fig.~\ref{fig_mose2wse2}, in Fig.~\ref{fig_MoS2mose2} we present results for the exciton energy for (a) different layer separations and (b) dielectric media of the spacer, but now for the MoS$_2$-MoSe$_2$ heterostructure. By comparing Figs.\ref{fig_mose2wse2} and \ref{fig_MoS2mose2}, one observes a similar overall behavior for the interlayer and intralayer excitons, owing to the screened interaction and the geometrical disposition of the heterostructure, showing qualitative physical trends that are independent of the TMD layers composition. Unlike the MoSe$_2$-WSe$_2$ case [see Fig.~\ref{fig_mose2wse2}(a)], for the MoS$_2$-MoSe$_2$ case, the lowest exciton energy for small layer separation is the interlayer IX$_1$, as emphasized in the inset of Fig.~\ref{fig_MoS2mose2}(a). As seen in Fig.~\ref{fig_MoS2mose2}(b), the dielectric media allows tuning both interlayer and intralayer exciton states, lowering their frequencies as larger the dielectric constant, exhibiting a more pronounced effect on the interlayer case.

Finally, we explore the spatial distribution of the exciton wave function (see Appendices~\ref{app_WF} and \ref{App_comparisonwithothermethods} for the analytical formulation of the configuration space wave function and the comparison of the assumed methodology here with other theoretical methods). Figures~\ref{fig_IntralayerMap_InterlayerMap}(a,b) and \ref{fig_IntralayerMap_InterlayerMap}(c,d) show color maps of the intralayer and interlayer exciton wave functions by varying (a,c) the interlayer distance $d$ and (b,d) the dielectric constant $\epsilon_m$ of the spacer. Figure~\ref{fig_IntralayerMap_InterlayerMap}(a) depicts no pronounced change in the spatial distribution of the intralayer exciton wave function when changing the interlayer distance. This can be linked to the energetic negligible changes in the binding energy as shown by the very small energetic scale variation in Fig.~\ref{fig_mos2_heteros}(a). On the other hand, as already expected, since by changing the dielectric constant the electron-hole interaction should vary, Fig.~\ref{fig_IntralayerMap_InterlayerMap}(b) shows different spatial distributions of the intralayer exciton wave function when varying the dielectric constant of the spacer. The higher $\epsilon_m$ value the lower the electron-hole interaction and consequently the binding energy value becomes smaller [see Fig.~\ref{fig_mos2_heteros}(b)] and thus the exciton wave function spreads more, \textit{i.e.} increasing the exciton size. Figures~\ref{fig_IntralayerMap_InterlayerMap}(c,d) demonstrate that the interlayer exciton wave function is much more sensitive to changes in the layer separation [Fig.~\ref{fig_IntralayerMap_InterlayerMap}(c)] than the intralayer case [Fig.~\ref{fig_IntralayerMap_InterlayerMap}(a)]. This is to be expected because the Coulomb interaction for interlayer exciton gets weaker with the increase of the layer separation, leading to spreading out the in-plane wave function. From Figs.~\ref{fig_IntralayerMap_InterlayerMap}(c,d), one notices that the wave function covers a larger spatial region for the interlayer case compared to the intralayer case [Figs.~\ref{fig_IntralayerMap_InterlayerMap}(a,b)], for both cases of changing the layer separation (being up to $35$\,\AA\ in panel (c)) and the interlayer dielectric constant (being up to $50$\,\AA\ in panel (d)). 

\section{Conclusions}\label{sec.conclusions}

In summary, we have presented a theoretical framework based on an appropriate expansion for the excitonic wave function basis composed here of the Chebyshev polynomials to solve the excitonic Wannier equation for double-layer heterostructure formed by different TMDs separated by a dielectric spacer. The employed method showed a fast convergence and numerical reliability with a computationally cheap scheme, owing to the recursive relations of the Chebyshev polynomials and the Chawla-Kumar decomposition that allowed us to integrate out the infrared divergence of the electron-hole interaction.

Based on the mentioned theoretical formalism, we explored the excitonic spectrum for intralayer and interlayer exciton configurations and its tunability through dielectric engineering, which arises from the screened Coulomb interaction. We reported that there is a robustness of the intralayer state with respect to the layer separation, while the interlayer exciton energy increases due to the binding energy sensitiveness to layer separation. By changing the dielectric media, the intralayer exciton energy decreases, although not as sharply as the interlayer exciton, which has the weakest binding for a large dielectric constant. Moreover, we also have obtained corrections to the bandgap using the semiconductor Bloch equations formalism, which enables us to understand how to layer separation and dielectric media affect the exciton energy. Our findings showed that even the energetic ordering relative to the intralayer and interlayer excitons can be modified by changes in the layer separation and in the dielectric constant of the spacer. Therefore, by dielectric engineering of the surrounding environment, we showed that the excitonic properties in double-layer van der Waals materials can be modified, enabling a bandgap control that suits different technological applications.

We hope that our theoretical framework and results based on Chebyshev's polynomial basis for Wannier excitonic complexes will prove useful for the exploration of optoelectronics properties in different van der Waals materials with a layer-by-layer stacking and surrounding environment controlling, and moreover being a simple and efficient tool for explaining cutting edge experiments in double layer 2D semiconductors, such as nonlinear optical susceptibilities.

\section*{Acknowledgments} 

This work is a part of the project INCT-FNA proc. No. 464898/2014-5. The work of M.~R.~H. was supported by the National Science Foundation under Grant No. NSF-PHY-2000029 with Central State University. K.~M. acknowledges a Ph.D. scholarship from the Brazilian agency CNPq (Conselho Nacional de Desenvolvimento Cient\'ifico e Tecnol\'ogico). K.~M., T.~F., A.~J.~C., and D.~R.~C. were supported by CNPq Grant No.~400789/2019-0, 308486/2015-3, 315408/2021-9, and 313211/2021-3, respectively. A.~J.~C. and T.~F. acknowledge Funda\c{c}\~ao de Amparo \`a Pesquisa do Estado de S\~ao Paulo (FAPESP) under Grant No.~2022/08086-0 and Thematic Projects 2017/05660-0 and 2019/07767-1, respectively. T.~A.~S.~P. are kindly thankful to CAPES (Coordenação de Aperfeiçoamento de Pessoal de Nível Superior) for financial support to the graduation course in Physics of the Federal University of Mato Grosso.

\begin{appendix}
 
\section{RK potential in a heterostructure} \label{subsec.poisson} 

In order to derive the RK potential for the chosen heterostructures, the Poisson equation has to be solved considering  three dielectric regions separated by two layers located at $z=0$ and $z=-d$ (see Fig.~\ref{Fig1}). Each layer has a polarization coefficient denoted by $r_1$ and $r_2$, respectively. Considering a charge $Q_1$ at $z=0$, we look for the potential distribution. The presence of a charge at the uppermost layer will induce a charge density $\rho_{ind}(\Vec{r})$ due to polarization. Therefore, the equation which we must solve is
\begin{gather}
-\nabla^2 \phi(\Vec{r})=\frac{1}{\epsilon_{0}}\rho(\Vec{r}). \label{eq:a1}
\end{gather}
Replacing the charge density $\rho(\Vec{r})$, one gets
\begin{equation}
-\nabla^2 \phi(\Vec{r})=\frac{1}{\epsilon_{0}}\left(Q_{1}\delta(\Vec{r})+ \rho_{ind}(\Vec{r})\right)\,.
\label{poisson}
\end{equation} 

The induced charge density term is
\begin{gather}
\rho_{ind}= \sigma_{1}\delta(z=0)+\sigma_{2}\delta(z+d)-\Vec{\nabla}\cdot\Vec{P}\, , \label{eq:a3}
\end{gather}
where $\Vec{P}$ is the medium polarization. If we consider that the medium polarization is linear, we can write the last term of Eq.~\eqref{eq:a3} as
\begin{gather}
\Vec{\nabla}\cdot\Vec{P}=\epsilon_{0}\chi_{i} \Vec{\nabla}\cdot{\Vec{E}}=-\epsilon_{0}\chi_{i} \nabla^2 \phi(\Vec{r})\,,
\label{eq:a4}
\end{gather}
which leads to the following partial differential equation
\bea\label{eq de poisson}
-\hspace{-0.1cm}\nabla^2\hspace{-0.1cm} \phi(\Vec{r}) \hspace{-0.1cm} = \hspace{-0.125cm}\frac{1}{\epsilon_{0}}\hspace{-0.15cm}\left[Q_{1}\delta(\vec r) \hspace{-0.1cm}+\hspace{-0.1cm} \sigma_{1}\delta(z) \hspace{-0.1cm}+\hspace{-0.1cm} \sigma_{2}\delta(\hspace{-0.05cm}z\hspace{-0.1cm}+\hspace{-0.1cm}d\hspace{-0.05cm}) \hspace{-0.1cm}+\hspace{-0.1cm} \epsilon_{0}\chi_{i}\nabla^2\hspace{-0.1cm}\phi(\Vec{r})\right]. 
\eea
Next, we apply a planar Fourier transform and rearrange Eq.~\eqref{eq de poisson}, which yields for $z>0$ to
 \begin{gather}\label{reg1}
 (1+\chi_{1})\left(q^2-\dfrac{\partial^2 }{\partial z^2}\right)\Phi(\Vec{q},z)=0,
 \end{gather}
where $\vec{q}$ denotes the planar Fourier components. A possible solution for Eq.~\eqref{reg1} is
 \begin{gather}\label{eq:a7}
\Phi(\Vec{q},z)=Ae^{-qz}+A'e^{qz},
 \end{gather}
and by noting that in the limit of large $z$ the potential should tend to zero, resulting to
\begin{gather}\label{eq:a8}
\Phi(\Vec{q},z)=Ae^{-qz}\,, \, z>0.
\end{gather}
Performing a similar procedure for the surrounded regions associated with the spacer and the substrate, we obtain, respectively
\begin{subequations}
    \begin{align}
    &\Phi(\Vec{q},z)=B\sinh{qz}+C\cosh{qz}\,,\, -d<z<0 \, ,\label{eq:a9} \\
    &\Phi(\Vec{q},z)=De^{qz}\,,  \, -d<z \, .\label{eq:a10}
    \end{align}
\end{subequations}    
Using the continuity of the potential, let us now rearrange Eq.~\eqref{eq de poisson} and integrate it around each of the layers, leading to a system of equations that allows us to determine the coefficients of the potential. Thus, rearranging Eq.~\eqref{eq de poisson}, we obtain
\begin{align}\label{eq:a11}
\hspace{-0.1cm}-\hspace{-0.05cm}(\hspace{-0.05cm}1 \hspace{-0.05cm}+\hspace{-0.05cm} \chi_{i} \hspace{-0.05cm})\hspace{-0.05cm} \nabla^2 \hspace{-0.05cm}\phi(\Vec{r}) \hspace{-0.05cm}=\hspace{-0.05cm} \frac{1}{\epsilon_{0}} \hspace{-0.05cm}\left[\hspace{-0.05cm} Q_{1}\delta(\vec r) \hspace{-0.05cm}+\hspace{-0.05cm} \sigma_{1}\delta(\Vec{z}) \hspace{-0.05cm}+\hspace{-0.05cm} \sigma_{2}\delta(z \hspace{-0.05cm}+\hspace{-0.05cm} d )\hspace{-0.05cm}\right],
\end{align}
and integrating around $z=0$, we get
\begin{multline}\label{eq:a12}
\int_{-\delta}^{+\delta}dz ~\epsilon_{i}\left(q^2-\dfrac{\partial^2}{\partial z^2}\right)\Phi(\Vec{q},z)=-\epsilon_{1}\left(\dfrac{\partial\Phi(\Vec{q},z)}{\partial z }\right)_{z=\delta}\\+\epsilon_{2}\left(\dfrac{\partial\Phi(\Vec{q},z)}{\partial z }\right)_{z=-\delta}=\frac{Q_{1}}{\epsilon_{0}}+\frac{\Sigma_{1}}{\epsilon_{0}}\, .
\end{multline}
Next, by evaluating the derivatives and taking the limit $\delta \to 0$, we arrive at
\begin{gather}\label{eq:a13}
\epsilon_{1}qA+\epsilon_{2}qB=\frac{Q_{1}}{\epsilon_{0}}+\Sigma_1{\epsilon_{0}}.
\end{gather}
The planar Fourier transform of $\sigma_1$ and $\Sigma_{1}$ can be found by using the in-plane polarization
\begin{gather}\label{eq:a14}
\sigma_{1}=-\Vec{\nabla}\cdot \Vec{P}_{\|}=-r_{1}\epsilon_{0}\left[\nabla^2 \phi(\Vec{r})\right]_{\|} \, ,
\end{gather}
which leads to
\begin{gather}\label{eq:a15}
\Sigma_{1}=-r_{1}\epsilon_{0}q^2\Phi(\Vec{q},z=0)=-r_{1}\epsilon_{0}q^2A\, .
\end{gather}
Replacing Eq.~\eqref{eq:a15} into Eq.~\eqref{eq:a13}, one gets one of the equations to obtain the coefficients $A$, $B$, $C$, and $D$ [see below Eq.~\eqref{sistema.a}]. Moreover, due to the continuity of the potential at the interface at $z=-d$, using Eqs.~\eqref{eq:a9}, \eqref{eq:a10}, and \eqref{eq:a11}, and also by taking the limit such that $\delta \to 0$, noting that $A=C$, one obtains the other two equations [Eqs.~\eqref{sistema.b} and \eqref{sistema.c}] of the system of equations
\begin{subequations}
    \begin{align}
        (\epsilon_{1}q+r_{1}q^2)A +\epsilon_{2}qB & = \frac{Q_{1}}{\epsilon_{0}}, \label{sistema.a} \\
        -B\sinh(qd)+A\cosh(qd) &= De^{-qd}, \label{sistema.b} \\
        \epsilon_{2}\left[B\cosh(qd)-A\sinh(qd)\right] &=(\epsilon_{3}+r_{2}q)De^{-qd}. \label{sistema.c}
    \end{align}
\end{subequations}
Using Eq.~\eqref{sistema.c}, we can write
\begin{gather}\label{eq:a17}
  De^{-qd}=\epsilon_{2}\frac{B\cosh(qd)-A\sinh(qd)}{\epsilon_{3}+r_{2}q},
\end{gather}
which in turn implies that only $A$ and $B$ are relevant. By defining the function $G_{j}(q)$ as
\begin{gather}\label{eq:a18}
  G_{j}(q)=\frac{\cosh(qd)(\epsilon_{3}+r_{j}q)+\epsilon_{2}\sinh(qd)}{\epsilon_{2}\cosh(qd)+\sinh(qd)(\epsilon_{3}+r_{j}q)},
\end{gather} 
the solution of the system of equations [\eqref{sistema.a}-\eqref{sistema.c}] for $A$ and $B$ results in
\begin{subequations}
    \begin{align}
        A &= \frac{-Q_{1}}{q\epsilon_{0}\left[\epsilon_{1}+r_{1}q+\epsilon_{2}G_{2}(q)\right]} , \label{eq:a19.a}\\
        B &= G_{2}(q)\frac{Q_{1}}{q\epsilon_{0}\left[\epsilon_{1}+r_{1}q+\epsilon_{2}G_{2}(q)\right]} . \label{eq:a19.b}
    \end{align}
\end{subequations}
The potential in momentum space is then given by
\be \label{eq:a20}
\Phi(\Vec{q},z) \hspace{-0.1cm}=\hspace{-0.1cm}
\begin{cases}
Ae^{q(z+d)}\left[\cosh(qz)+G_{2}(q)\sinh(qz)\right] ; z<-d,\\
A\left[\cosh(qz)-G_{2}(q)\sinh(qz)\right] ; -d<z<0,\\
Ae^{-qz} ; z>0. 
\end{cases}
\ee
Since we are particularly interested in the intralayer and interlayer effects, we can write explicitly, using Eqs.~\eqref{eq:a18}, \eqref{eq:a19.a}, \eqref{eq:a19.b}, \eqref{eq:a20}, and by also doing some relabeling, the following expressions
\begin{subequations}
    \begin{align}
        V_{ii}(q) &= \frac{-e^2}{q\epsilon_{0}\left[\epsilon_{1}+r_{i}q+\epsilon_{2}G_j(q)\right]}, \label{Viipot}\\
        V_{i,j\neq i}(q) &= \frac{e^2\left[\cosh(qd)-G_j(q)\sinh(qd) \right]}{q\epsilon_0\left[\varepsilon_1+r_i q+\varepsilon_2G_j(q) \right]}, \label{Vijpot}
    \end{align}
\end{subequations}
where $V_{ii}(q)$ and $V_{i,j\neq i}(q)$ are the intralayer  and the interlayer potentials, respectively. A interesting property of the $G_{j}(q)$ function [Eq.~\eqref{eq:a18}] is that
\begin{gather}\label{eq:a23}
\lim_{d\to \infty} G_{j}(q)= \lim_{d\to \infty}\ \frac{e^{qd}(\epsilon_{3}+r_{j}q)+\epsilon_{2}e^{qd}}{\epsilon_{2}e^{qd}+e^{qd}(\epsilon_{3}+r_{j}q)}=1.
\end{gather}
Using this result in Eq.~\eqref{Viipot}, we arrive at a fairly familiar result
\begin{gather}
V_{RK}(q)=\frac{-e^2}{q\epsilon_{0}(1+\bar{r}_{1}q)}, \label{eq_RK}
\end{gather}
where $\bar{r}_{1}=r_{1}/(\epsilon_{1}+\epsilon_{2})$. Equation~\eqref{eq_RK} is the RK potential in momentum space. A comparison between the derived intralayer [Eq.~\eqref{Viipot}] and interlayer [Eq.~\eqref{Vijpot}] potentials and the Coulomb potential is shown in Fig.~\ref{Fig2}.

\section{$\lambda$ recurrence relations} \label{app_CK}

We define
\begin{gather}
  \lambda_{i}(u) = \int_{-1}^{1}du^{\prime} \frac{T_{i}(u^{\prime})}{u-u^{\prime}},
\end{gather}
that obeys the following relations
\begin{gather}
  \lambda_{0}(u) = \ln{\left\lvert\frac{1+u}{1-u}\right\rvert}, \\
  \lambda_{1}(u) = -2 +u\lambda_{0}(u), \\
  \lambda_{k+1}(u) -2u\lambda_{k}(u) + \lambda_{k-1}(u) = 2\frac{\left[1+\cos(k\pi)\right]}{k^2-1}.
\end{gather}
Such recurrence relations and definitions are used in the analytic solution of $I_{n,l}$ in Eq.~\eqref{eq_In} in the Chebyshev method's Section \ref{subsec.tcheb}.

\section{Fitting procedure} \label{app_fitt} 

Our goal is to describe the electrostatic effects due to the geometry presented in Fig.~\ref{fig_band_schm}, starting from the exciton binding energy and bandgap of suspended monolayer samples. For this, we consider the experimental A exciton energy $E_A$ measured for suspended samples \cite{Klots2014,Xie2021,Harats2020,Aslan_2022}, the bandgap $\Delta_K$ calculated in Ref.~[\onlinecite{Zhang_2016}], and the SOC splitting of Ref.~[\onlinecite{Kormanyos_2015}]. The electron and hole of a bright exciton come from bands with the same spin and valley indexes, thus, for negative SOC$_\mathrm{CB}$ (see Fig. \ref{fig_band_schm}), the exciton binding energy is blue-shifted for the same magnitude. 

First, we obtain the screening length $r_0$ fitting the value of the binding energy of Table~\ref{tab_fit} for each MX$_2$ by solving the Wannier equation \eqref{eq_sch_partial} with the RK potential \eqref{eq_RK}. With this value of $r_0$, we solve the gap equation \eqref{eq_dressed}, also considering the RK potential, to obtain the ``bare'' transition energy $\Delta_{s\tau,0}$, which gives the transition energy calculated by Ref.~[\onlinecite{Zhang_2016}].

\begin{table}
\caption{Effective masses, screening factor $r_{0}$, and the bandgap of each material. The masses are obtained from Ref.~[\onlinecite{Kormanyos_2015}] and the screening factors are obtained via a fitting procedure.}
 \begin{tabular}{l| c c c c c c} \hline \hline
 Materials & $m_{e}$ \cite{Kormanyos_2015} & $m_{h}$ \cite{Kormanyos_2015} & $r_0$ & $r_{0}$\cite{r0dft}& $\Delta_K$\cite{Zhang_2016}(eV) & $E_\mathrm{b}$(meV) \\ \hline
 MoS$_2$ & $0.47$ & $0.54$ & $27.04$\AA & $23.45$\AA & $2.71$ & $-753.0$\cite{Klots2014} \\ 
 MoSe$_2$ & $0.58$  & $0.6$ & $35.34$\AA & $26.13$\AA & $2.37$ & $-711.7$\cite{Xie2021} \\
 WS$_2$& $0.27$ & $0.36$ & $20.85$\AA & $16.59$\AA & $2.91$ & $-900.0$\cite{Harats2020} \\
 WSe$_2$ & $0.29$ & $0.36$ & $21.80$ \AA & $20.09$\AA & $2.57$ & $-890.0$\cite{Aslan_2022} \\ \hline
\end{tabular} \label{tab_fit}
\end{table}

\section{Configuration space wave function}\label{app_WF}

Once Eq.~\eqref{eq_final} is solved, we can obtain the wave function in configuration space using the Fourier transform
\be
\psi_{n,\ell}(\br)= \int d^2\bp \, e^{i\bp \cdot \br} \psi_\ell(p) e^{i\ell \phi^\prime}.
\ee
By implementing the angular integration, we have that
\bea
\psi_{n,\ell}(r,\phi) &=& \frac{2}{\xi^2}e^{i\ell\phi} \sum_n c_{n,\ell}\int_{-1}^1 du \frac{1+u}{(1-u)^3}  \cr &\times& 
J_\ell\left(\frac{r}{\xi}\frac{1+u}{1-u} \right) f(u)T_n(u),
\eea
with $J_\ell$ being the Bessel function of order $\ell$. A computationally convenient choice for $f(u)$ is given by
\be
f(u)\equiv \frac{(1-u)^3}{1+u},
\ee
since it demonstrated a fast convergence. To understand the assumed procedure, let's exemplify with the calculation of the following quantity $F(q)$ of interest
\be
\langle F\rangle=\int dq q F(q) \psi_{n_1}(q)...\psi_{n_N}(q).
\ee
To do this, first, we write the above equation in terms of $u$
\bea \label{eq.example}
\langle F\rangle &=&  \int du F(q(u)) \left[\frac{1+u}{(1-u)^3} f(u) \right]^{n_N} \cr &\times& \sum_{j_1...j_{n_N}}c^{n_1}_{j_1}T_{j_1}(u)...c^{n_N}_{j_{n_N}}T_{j_{n_N}}(u).
\eea
The next step is to write the integrand of Eq.~\eqref{eq.example} in terms of a single Chebyshev expansion
\begin{align}
& F(q(u)) \left[\frac{(1-u)^3}{1+u} \right]^{n_N-1} \hspace{-0.25cm}\times\hspace{-0.25cm} \sum_{j_1...j_{n_N}}\hspace{-0.25cm} c^{n_1}_{j_1}T_{j_1}(u)...c^{n_N}_{j_{n_N}}T_{j_{n_N}}(u) \nonumber \\ 
&= \sum_k b_k T_k(u).
\end{align}
To do this, we use the procedure of convolution explained in Appendix~\ref{app_CK}. After that, we can use the Clenshaw-Curtis to obtain
\be
\langle F\rangle=\sum_{k=0}^\infty\frac{2b_{2k}}{1-(2k)^2}.
\ee

\section{Comparison with other methods}
\label{App_comparisonwithothermethods}

In order to corroborate our obtained results in Sec.~\ref{sec_results}, it is important to compare the Chebyshev method with a different method for solving the integral equation Eq.~\eqref{eq_sch_partial}. For this purpose, let's compare the method discussed in the paper with the more traditional quadrature method: the Gauss-Legendre quadrature. Let's rewrite Eq.~\eqref{eq_sch_partial} as
\begin{gather}
  \psi_{\ell}(p) = \frac{1}{E-E_{p}}\int_{0}^{\infty}\frac{dp'}{2\pi} p' V_{\ell}(p,p') \psi_{\ell}(p'),
\end{gather}
by rewriting the integration as a Gauss-Legendre quadrature and applying a hyperbolic mapping, we have
\begin{gather}
  \psi_{\ell}(p) = \frac{1}{E-E_{p}}\sum_{i}\frac{\omega_{i}(1+x_{i})V_{\ell}(p,x_{i})\psi_{\ell}(x_{i})}{\pi(1-x_{i})^3},
\end{gather}
which is a system of equations in which we search for unit eigenvalues with different input energies $E$. Results obtained via the Gauss-Legendre quadrature for the exciton ground state binding energy of MoS$_2$ for different numbers of mesh points for the radial momenta and fixed angular mesh points are shown in Table~\ref{tab_comparisontab}. From Table~\ref{tab_comparisontab}, one can see that the Chebyshev method, whose resulting value is $E_b = -753.0$ meV, agrees with the interpolated Nystr\"om method very well, which is a more computationally demanding method and for a quadratic extrapolation ($N_p\rightarrow\infty$) gives $E_b = -753.1$ meV, \textit{i.e.} showing an energetic difference between the methods of $0.1$ meV.

\begin{table}[t]
    \centering
    \caption{The convergence of exciton ground state binding energy as a function of the number of radial momenta mesh points $N_p$, obtained with RK potential for the MoS$_2$ from the Gauss-Legendre quadrature method. The number of angular mesh points is $61$. The exciton binding energy of $-753.1\,$meV for $N_p \to \infty$ is obtained with a quadratic extrapolation, while the Chebyshev Method yields a binding energy of $-753.0\,$meV.}
     \begin{tabular}{l| c } \hline \hline
     $N_{p}$ & $E_{b}$ (meV) \\ \hline
     $300$ & $-788.3$ \\ 
     $400$ & $-778.8$ \\
     $500$ & $-773.3$ \\
     $600$ & $-769.5$ \\ 
     $700$ & $-768.2$ \\
     $800$ & $-765.4$ \\
     $900$ & $-764.1$ \\
     $1000$ & $-762.9$ \\ \hline
     $N_p \to \infty$ & $-753.1$ \\ \hline
    \end{tabular}
    \label{tab_comparisontab}
\end{table}

\begin{figure}[t]
\centering
\includegraphics[width=1\columnwidth]{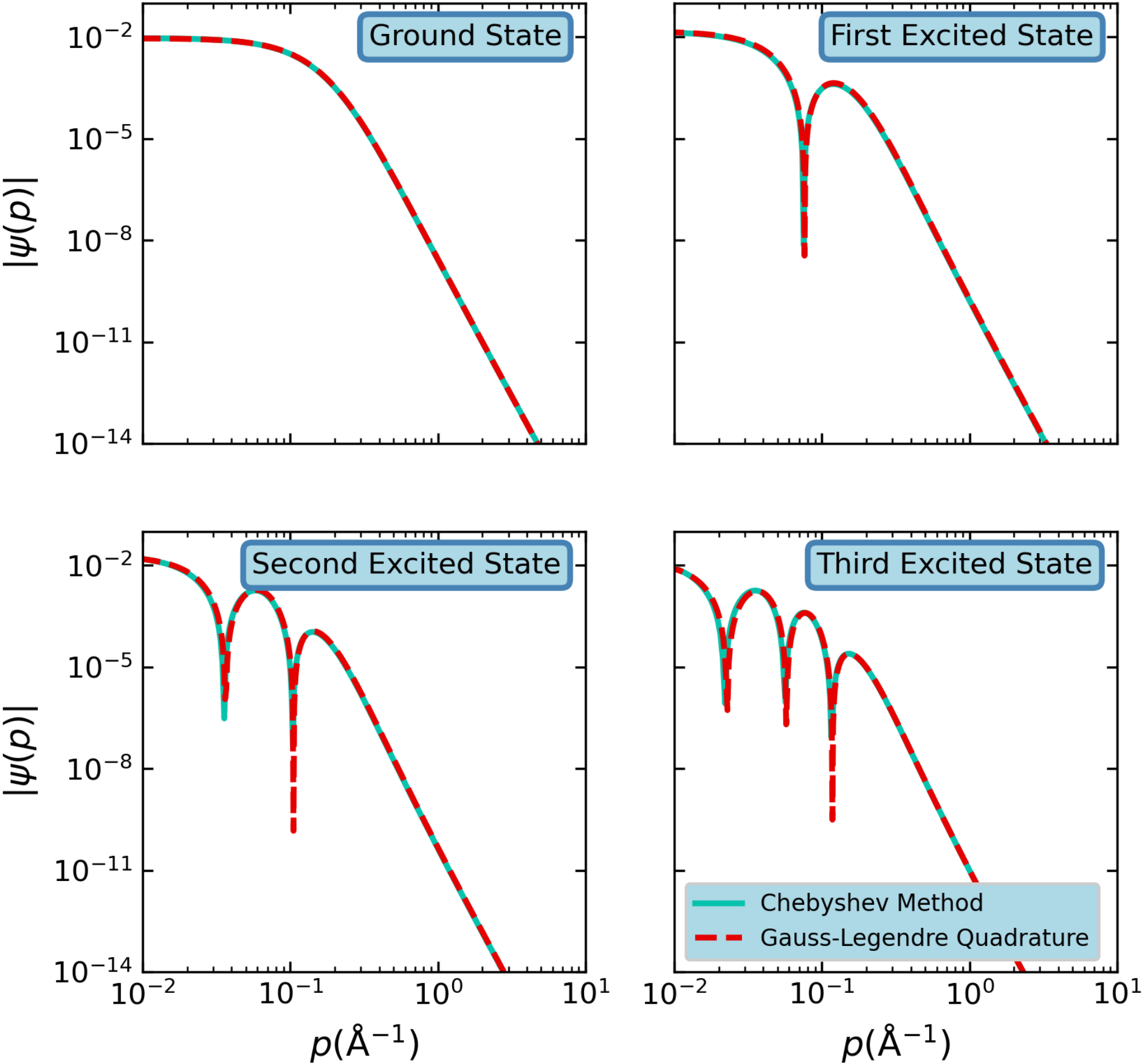}
\caption{(Color online) Comparison of wave functions obtained via (dashed red curves) the Gauss-Legendre quadrature method and (solid cyan curves) the Chebyshev method for the first four states of the MoS$_2$ exciton using the RK potential. $\psi_i(p)$ is the $i$-th excited state for the $s$-wave. Note that the $y$-axis is in log scale and the number of peaks represents the number of nodes in the excitonic wave function.}
\label{fig_comparison}
\end{figure}

To further validate our method, we also compare the wave functions for the first four states, \textit{i.e.} ground, first excited, second excited, and third excited states, obtained via the Chebyshev method (solid cyan curves) and the Gauss-Legendre quadrature (dashed red curves) in Fig.~\ref{fig_comparison}, assuming the RK potential for the interlayer electron-hole interaction. Figure~\ref{fig_comparison} shows that both methods are very reliable and generated similar quantitative and qualitative results. However, the Chebyshev method exhibits some oscillations for large momenta, where the wave function is in the order of $10^{-22}$.

Another good comparison for the binding energy value could be achieved with variational-like methods such as the one by Griffin, Hill, and Wheeler (GHW)\cite{GCMs, Mohallem1986}. Here, we consider a basis with a set of parameters $\zeta$ and calculate the secular equation generated by the inner product with the Hamiltonian in real space. The basis chosen is
\begin{gather}
  \psi_{n,\ell}(\Vec{r}) = A_{n} r^{\lvert \ell \rvert} e^{i\ell\phi}\sum_{j}c^{n}_{j}e^{-\zeta_j r},
\end{gather}
which yields
\begin{gather}
  \sum_{j}\left[H(\zeta_i,\zeta_j) - S(\zeta_i,\zeta_j) E_{n}\right]c_{j}^{n} = 0, 
\end{gather}
where
\begin{subequations}
\begin{align}
  H(\zeta_i,\zeta_j) &= \int d\br\psi^{*}_{n,\ell}(\Vec{r}) H \psi_{n,\ell}(\Vec{r}), \\
  S(\zeta_i,\zeta_j) &= \int d\br\psi^{*}_{n,\ell}(\Vec{r}) \psi_{n,\ell}(\Vec{r}).
\end{align}
\end{subequations}
The set of values for the parameter $\zeta$ is chosen in a logarithmic grid, such as $\Omega = \Gamma^{-1}\ln \zeta $. Here, we take $\Gamma = 5$ and set the interval $[-2,2]$. The number of points by which we subdivide the interval is obtained by trial and error, which yields $N = 48$. By choosing this set of parameters and grid, we arrive at a binding energy of $E_{b} = 752$\,meV, which shows a good agreement with our Chebyshev results.

\end{appendix}

\bibliography{references}

\end{document}